\documentclass[12pt]{article}
\usepackage{harvard}
\usepackage{bibtexlogo}
\usepackage{times}
\usepackage{color}
\usepackage{graphicx}

\pdfoutput=1

\def\farcs{\hbox{$.\!\!^{\prime\prime}$}}

\begin{document}
\bibliographystyle{agsm}

\title{How selection and weighting of astrometric
observations influence the impact probability. \\ Asteroid
(99942)~Apophis case.}
\author{Ma\l gorzata Kr\'olikowska$^1$, Grzegorz Sitarski$^1$ and Andrzej M. So\l tan$^{2}$\\
\footnotesize{$^1$Space Research Centre of the Polish Academy of Sciences,
Bartycka 18A, 00-716 Warsaw, Poland }\\
\footnotesize{$^2$Nicolaus Copernicus Astronomical Center, Bartycka 18, 00-716
Warsaw , Poland}}


\maketitle

\abstract{The aim is to show that in case of low probability of
asteroid collision with Earth, the appropriate selection and
weighing of the data are crucial for the impact investigation, and
to analyze the impact possibilities using extensive numerical
simulations. By means of the Monte Carlo special method a large
number of ``clone'' orbits have been generated. A full range of
orbital elements in the 6-dimensional parameter space, e.g. in the
entire confidence region allowed by the observational material has
been examined. On the basis of 1000 astrometric observations of
(99942)~Apophis, the best solution for the geocentric encounter
distance of $6.065\pm 0.081$\,$R_{\oplus}$ (without perturbations
from asteroids) or $6.064\pm 0.095$\,$R_{\oplus}$ (including
perturbations by four largest asteroids) were derived for the close
encounter with the Earth on April 13, 2029. The present
uncertainties allow for the special configurations (``keyholes'')
during these encounter which may lead to the very close encounters
in the future approaches of Apophis. Two groups of keyholes are
connected with the close encounter with the Earth in 2036 (within
the minimal distance of $5.7736-5.7763$\,$R_{\oplus}$ on April 13,
2029) and 2037 (within the minimal distance of
$6.3359-6.3488$\,$R_{\oplus}$). The nominal orbits for our most
accurate models run almost exactly in the middle between these two
impact keyhole groups. A very small keyhole for the impact in 2076
has been found between these groups at the minimal distance of
$5.97347$\,$R_{\oplus}$. This keyhole is close to the nominal orbit.
The present observations are not sufficiently accurate to eliminate
definitely the possibility of impact with the Earth in 2036 and in
many years following this year. It is shown that the available seven
radar measurements are not crucial at present for the nominal orbit
determination. }


\section{Introduction}

The discovery of potentially dangerous asteroid often starts the
alarm of the world community because of its possible collision with
Earth in the foreseeable future. Fortunately, so far this potential
risk of collision decreases as more observations are successively
collected. Up to date the impact probability estimates of known
Potentially Hazardous Asteroids (PHA; at the beginning of 2009 there
were more than 1000 such objects) are at the outmost in the range of
$10^{-4}-10^{-5}.$\footnote{At the moment of writing, on the top of
the list: 'Objects Not Recently Observed' (Sentry Risk Table, NASA)
are (101955) 1999~RQ$_{36}$ with cumulative probability of $7.1\cdot
10^{-4}$ and 2007~VK$_{184}$ with cumulative probability of
$3.4\cdot 10^{-4}$.} The main aim of this paper is to show that in
the case of so low probabilities, the appropriate selection and
weighing of the data are crucial for the impact investigation. To
illustrate this question, we scrupulously examined the observational
material and made extensive Monte Carlo analysis of the future
encounters with the Earth by the asteroid (99942) Apophis.

This is a potentially dangerous object since it has a large size
(diameter $270\pm 60$ meters, \citeasnoun{delbo:2007}) and future
collision possibilities have not been definitively solved yet.
Additionally, Apophis will not be observable until 2011
\cite{Ches:2006}. The observational data collected in the months of
March 2004 and August 2006 consist of $1000$ optical and 7 radar
measurements. In the present paper we concentrate on the astrometric
observations alone and show that the adequate selection and
weighting procedures applied to these observations provide the
nominal orbit with the same accuracy as the estimates found in the
literature, based on the astrometric {\it and} the radar data. This
is due to a large disproportion between the number of optical and
radar measurements. If the number of radar observations were
significantly greater or the radar data were outside the optical
interval of data, the situation would be different.

We investigate the Apophis motion as a pure ballistic problem. Thus,
we ignore the non-gravitational (NG) effects. Obviously, to describe
accurately the asteroid orbit, these effects should be included. The
problem of the NG effects is widely discussed by
\citeasnoun{gio:2008}. They show, that the present data are
absolutely insufficient to construct any decent model of these
effects. Thus, we are unable to predict precisely the Apophis
trajectory in the distant future. The purely gravitational
computations have been perform to show the potential Apophis
behaviour, especially the wider -- than given in the literature --
keyhole ranges in 2036 and 2037 resulting from our full 6D Monte
Carlo method.

Some details of the Apophis story are worthy of notice. Asteroid was
discovered by Tucker, Tholen and Bernardi at Kitt Peak (Arizona) on
June 19, 2004. Unfortunately, the object was lost until December 18,
when it was rediscovered by Garradd from Siding Spring in Australia.
On the basis of six month of observations Apophis was recognized as
a potentially hazardous asteroid with non-zero impact probability in
2029. However, substantial astrometric errors in the original June
observations were quickly revealed \cite{Ches:2006}. After
remeasurements done by Tholen the impact probability was assessed at
about $0.6$\,\% and during the next days was systematically
increasing reaching a peak of $2.7$\,\% at the end of December. The
pre-discovery observations from March 2004, reported by the
Spacewatch survey at the end of December, eliminated any possibility
of an impact in 2029. Calculations based on observations from the
March through December have shown that the asteroid will pass near
Earth on April 13, 2029 in the minimum distance of $10.1\pm
2.6$~R$_{\oplus}$ from the geocenter (R$_{\oplus}=6378$\,km).
Moreover, it turned out that this deep encounter with Earth in 2029
would imply resonant return encounters in the subsequent years that
could lead to several impact possibilities.

Later, the radar astrometry obtained in late January 2005 from the
Arecibo Observatory were reported to be inconsistent with this
prediction \cite{IAUC8477}. \citeasnoun{Gio:05} found that radar
data indicated a significantly closer approach of $5.6\pm
1.6$~R$_{\oplus}$. According to \citeasnoun{Ches:2006} the
discrepancy was explained by the systematic errors in the five
pre-discovery observations of March 2004 and the remeasurements of
these observations were done by Spacewatch team and Spahr from MPC
staff. The exciting story about changing the collision scenario of
Apophis during the December 2004 and January 2005 is described in
details by \citeasnoun{Sansaturio:08}.

\begin{table}
\caption{Minimal distance in April 2029 for the different
observational intervals. The weighting procedure was applied for
each case independently.} \label{tab:arcs} \vspace{0.10cm}
\begin{center}
\footnotesize{
\begin{tabular}{cccccc} \hline
 Solution &  Observational  & Number & Number     & rms         &  Minimal distance     \\
          &   interval      & of     &  of        &             &  on April 13, 2029    \\
          &                 & obs.   & residuals  &             &  [R$_{\oplus}$]       \\
\hline
 arc1 & 2004\,06\,19 -- 2004\,12\,27 & 264 &  520 & 0\farcs 339 &$48.56\pm 6.98$     \\
 arc2 & 2004\,03\,15 -- 2004\,12\,27 & 270 &  535 & 0\farcs 352 & $5.542\pm 0.475$   \\
 arc3 & 2004\,03\,15 -- 2005\,03\,26 & 892 & 1771 & 0\farcs 316 & $6.699\pm 0.267$   \\
 arc5 & 2004\,03\,15 -- 2006\,06\,02 & 994 & 1965 & 0\farcs 316 & $6.564\pm 0.156$   \\
  E   & 2004\,03\,15 -- 2006\,08\,16 &1000 & 1971 & 0\farcs 308 & $6.065\pm 0.081$   \\
 arc6 & 2004\,12\,18 -- 2006\,08\,16 & 988 & 1965 & 0\farcs 314 & $6.144\pm 0.078$   \\
 \hline
\end{tabular}}
\end{center}
\end{table}

According to \citeasnoun{gio:2008} the new Arecibo radar
observations of Apophis in August 2005 and May 2006 have increased
the close approach distance on April 13, 2029 to $5.86\pm
0.11$~R$_{\oplus}$ and $5.96\pm 0.09$~R$_{\oplus}$, respectively
(38\,000$\pm 580$\,km; closer than some geosynchronous communication
satellites).

Our Table~\ref{tab:arcs} serves as a comment to the Apophis {\it
varying} approaches to the Earth on April 13, 2029. We give there
minimal distance from the Earth derived by us for the six different
observational arcs based solely on the astrometric observations
(i.e. excluding the seven radar observations). It is worth to note
that the results based on the ``arc6'' in the Table~\ref{tab:arcs}
which use neither the recalculated a posteriori observations of
March and June 2004 nor the radar measurements, are similar to the
value derived by \citeasnoun{gio:2008} on the basis of all the
astrometric and radar data. It that after June 2004 there are no
inconsistencies between the radar and astrometric data.

Though the risk of a collision with the Earth or the Moon in 2029
has been eliminated, there remains still a very small possibility
that during the close encounter with Earth on April 13,~2029,
Apophis would pass through a ``gravitational keyhole'', a precise
region in space that would set up a future impact on April 13,
2036\footnote{The term keyhole is used here according to its
classical meaning introduced by \citeasnoun{Chodas:1999}. This term
may also be used to indicate a region on the target plane of the
first encounter leading (at the subsequent return) not necessarily
to the collision, but to a deep encounter (for more details see
\citeasnoun{val:2003}).}. Our numerical calculations show that
though the keyhole in 2029 for the 2036 impact is several times
larger than 400 meters given by many authors, the impact risk is
still extremely low.

In this paper we present details of selection and weighting of
Apophis observations and their effect on the best estimates of its
position during the close safe encounter with Earth in 2029 and the
possibility of impacts in 2036 and 2037. We are able to directly
determine the sample of impact orbits for each close encounter with
the Earth.

According to our impact calculations (Section~\ref{impactan}),
Apophis will hit Earth in 2036 only if it passes through a keyhole
on April 13, 2029, which is a roughly 4.6 kilometer wide region in
space lying within $5.7736-5.7744$\,R$_{\oplus}$ from the Earth's
geocenter. Another dangerous possibility is that Apophis will pass
through the second 6.4\,kilometer wide keyhole lying within
$6.3395-6.3405$\,R$_{\oplus}$ from the Earth's geocenter. The last
one leads to a collision in April 2037. We also determined a few
other extremely small keyholes leading to impacts after 2037. These
keyhole ranges were obtained using extensive Monte Carlo
simulations. A large samples of VAs in a full 6-dimensional
uncertainty region of orbital elements (or position-velocity region)
have been generated. Thus, the analysis has been constrained to a
pure ballistic problem. A similar approach has been applied by
\citeasnoun{gio:2008} who used the Monte Carlo method in the
six-dimensional position-velocity space. They examined the Apophis
positional uncertainty after 2029, but did not investigate the
Apophis impact orbits.

The equations of cometary motion was integrated numerically using
the recurrent power series method (Sitarski 1989, 2002) by taking
the perturbations by all the planets and by the Moon into account.
The perturbations from four largest asteroids (Ceres, Pallas, Vesta,
Hygiea) are included only for Model~E$^{\prime}$. It allows to
estimate the influence of these objects on the impact risk
probability. All numerical calculations presented here are based on
the Warsaw numerical ephemeris DE405/WAW of the Solar System,
consistent with a high accuracy with the JPL ephemeris DE405
\cite{Sit:02}. The positional observations of Apophis have been
taken from the NEODyS pages publicly available at
http://newton.dm.unipi.it/neodys/.

\section{Selection and weighting of astrometric observations}

Selecting and weighting the astrometric observations constitute a
crucial procedure for the asteroid orbit determination process.
Individual groups use different methods of the data preparation. For
example, in the Apophis case the researchers from the Jet Propulsion
Laboratory rejected about $26$\,\% of optical measurements. The
resulting rms from $738$ optical measurements and $7$ radar
observations is reduced to $0\farcs 352$.  Similarly,
\citeasnoun{gio:2008} who had rejected about $21$\,\% of optical
data and kept $7$ radar measurements, determined the nominal orbit
with the rms of $0\farcs 407$. One of the widely used method of the
data selection and weighting -- the 'global residual statistics' is
described by \citeasnoun{carpino:2003}. It is based on a global {\it
O-C} statistics of the optical astrometric observations collected
for about 17000 numbered asteroids. The `global weights' are then
used for an automatic orbital analysis of asteroids by many authors.
This method has been applied for Apophis by the Near Earth Objects -
Dynamic Site, where only $5$ of $1000$ astrometric observations were
rejected and the resulting rms is $0\farcs 302$ (the radar data were
also included). However, the inspection of details of the data
processing shows that for $68$\,\% of optical observations the
weights have attribute 'forced', what implies that `manual'
intervention has been applied to the majority of
observations\footnote{This is despite the information in the www
page that such data handling is rarely applied.}.

Similarly to \citeasnoun{carpino:2003} we use the objective
statistical method, however we treated the existing set of
observations of each individual asteroid as the unique one. In fact,
we used our method (still improving in details) since more than 20
years thus we give next only a brief description of the criteria
used for statistical data analyzing.

To investigate the influence of the data selection and weighting on
the existence of impact orbits, especially on the probability of the
Earth's impact, we have prepared the first two sets of observations
applying Bielicki's and Chauvenet's criterion for selection
procedure and treating all the data points as equivalent
observations (solutions A and B, respectively; see
Table~\ref{tab:cases}). Both criteria differ in the upper limit of
the accepted residuals, $\xi$, e.g. observed minus computed values
of right ascension, $\Delta \alpha \cdot \cos \delta$, and
declination, $\Delta \delta$. According to the Chauvenet's criterion
\cite{cha:1908} from the set of $N$ residuals, $\xi$, we should
discard all values of $\xi$ for which

$$\mid \xi \mid > \sigma \cdot K_{1/2}(N)$$
where $\sigma$ is a dispersion of $\xi$:

$$\sigma = \sqrt{\left ( \sum_k \xi _k2 \right )/N}$$
and $K_{1/2}(N)$ is the unknown upper limit of the integral of the
probability distribution, $\phi (\xi)$:

$$ \int _0^{K_{1/2}} \phi (x) {\rm d}x = 1-\frac{1}{2N}\,,$$
where $x = \xi / \sigma$.

According to this criterion the data point is rejected if the
probability of obtaining the particular deviation of residuals from
the mean value is less than $1/(2N)$. To determine this probability
the normal distribution of $\xi$ is assumed.

\noindent In the less restrictive Bielicki's criterion \cite{Bie72}
the data points are rejected if:

$$\mid \xi \mid > \xi _{K_B}= \sigma \cdot K_{1/2}(N)/(1-0.4769363\sqrt{N})\,.$$
It is taken into account here that the dispersion $\sigma$ itself is
a random variable.

We also used the Bessel criterion that is more restrictive than the
Chauvenet's criterion. The Bessel criterion rejects from the set of
$N$ residuals all the values of $\xi$ for which

$$\mid \xi \mid > \sigma \cdot K_{1}(N)\,,$$
where $K_{1}(N)$ is defined by:

$$ \int _0^{K_{1}} \phi (\xi) {\rm d}\xi = 1-\frac{1}{N}.$$

To reduce systematic errors in the observational material, such as
the bias associated with a site as a function of time, one should
consider specific procedures. In the present investigation we
divided the whole observational material into several time
subintervals according to the inertial structure of material (i.e.
according to existing gaps in observations).

\begin{figure}
\begin{center}
\bibliographystyle{agsm}
\includegraphics[width=8.0cm]{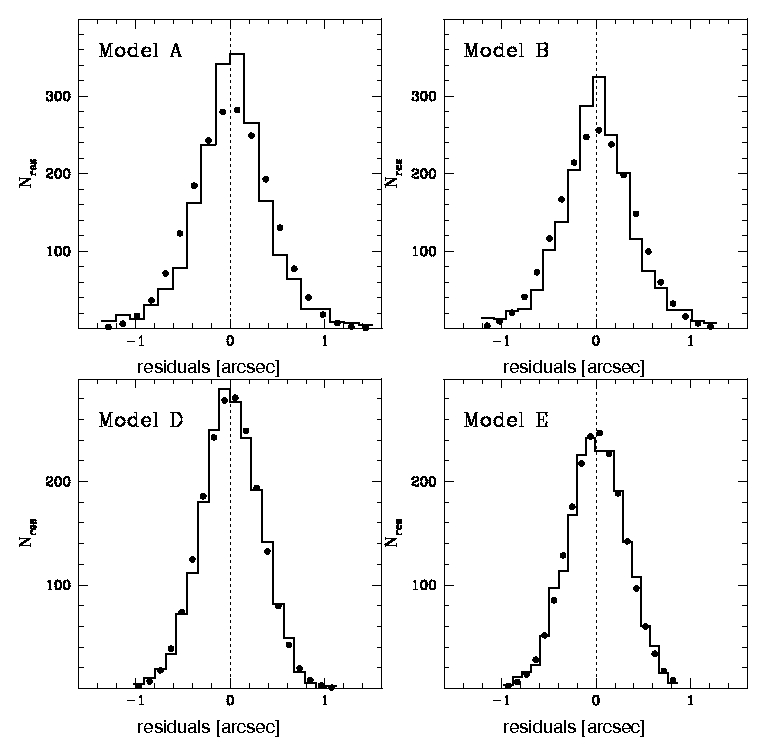}
\caption{The O-C distributions for the non-weighted data (upper
panel, Models A,B) and the weighted data (lower panel, Models D,E).
The best-fitting Gaussian distributions are shown by black dots. }
\label{fig:OCdist}
\end{center}
\end{figure}

The application of the Chauvenet's criterion to the Apophis data
resulted in the rejection of 14 more residuals than using the
Bielicki's criterion (Model~A~and~B in Table~\ref{tab:cases},
column~3).  Our selection method allows us to discard any ``bad''
residual in right ascension keeping  ``good'' residual in
declination, and vice versa. In the set of the Apophis observations
the Bessel criterion resulted in rejection of only a few residuals
more than the Chauvenet's criterion. To visualize the importance of
the data selection we have constructed the Model~C, in which we
arbitrarily removed all the residuals with {\it O-C} greater than
$0.6$\,arcsec. It is important to stress that ignoring some
statistically  acceptable data points (in this case about $29$\,\%
of all the observations) one can affect the data in statistically
unacceptable way. The fact that, due to the smallest rms value,
Model~C looks more attractive than Models~A and~B, cannot be used as
an argument favouring this model.

Next, two sets of data (solutions~D and~E) were handled by the
iterative procedure of selection and weighting the observations. At
the end of the iteration the computed weights were normalized to
unity for all the observers. The scheme of this procedure was
described in details by Bielicki and Sitarski (\citeyear{BieSit91}).
Solution~D is based on the Bielicki's criterion of selection while
the solution~E -- on Chauvenet's criterion. One can see from
Table~\ref{tab:cases} that weighting and selection procedure leads
to the significantly smaller mean residuals and restores more data
than the selection procedure alone.

After fitting the Gaussian model to the {\it O-C} distributions for
all five nominal orbits we concluded that distributions of residuals
for the two non-weighted models~A and B show some deviations from
the Gaussian model. These deviations can be described by kurtosis,
$K$ (related to the fourth moment of the distribution) and skewness,
$\gamma_1$ (related to the third moment). We use standard
definitions of both quantities: $K = \frac{\mu_4}{\sigma4} - 3\,$,
where $\mu_4$ is the fourth central moment, $\sigma$ is the standard
deviation, and $\gamma_1 = \frac{\mu_3}{\sigma3}\,,$ where, $\mu_3$
is the third central moment.

The values of kurtosis and skewness are given in
Table~\ref{tab:cases}. The amplitudes of skewness at about $-0.1$
for the {\it O-C} distributions in the case of the non-weighted data
(Model~A and B) indicate that these distributions are satisfactorily
symmetric. However, kurtosis for these samples are equal to $1.3$
and $0.9$, respectively. Thus, these distributions are leptokurtic
-- with a distinct peak at the mean as compared to the Gaussian
distribution (Fig.~\ref{fig:OCdist}). It means that the classical
assumption that the observation errors are distributed according to
the Gaussian probability density function, is not true in the
Apophis case.

\begin{table}
\caption{Orbital models for Apophis. Solutions A, B, C and  D,
E/E$^\prime$ differ in the assumed criterion of selection. The data
for the first three models were processed without weighting, while
for the latter two -- with weighting. In column~5 threshold values
of rms for the confidence level $\alpha = 0.99$ are given (see
Sect.~\ref{cloning} for details).} \label{tab:cases}
\begin{center}
{\setlength{\tabcolsep}{1.3mm}\footnotesize{
\begin{tabular}{cccccccccc} \hline
 Solution &  Observational  & Number of  & rms    & rms$_{99}$        & $K$    & $\gamma_1$ & Minimal distance   & \multicolumn{2}{c}{Impact}      \\
          &   interval      & residuals  &        &                   &        &            & on April 13, 2029  & \multicolumn{2}{c}{probability } \\
          &                 &            &        &                   &        &            & [R$_{\oplus}$]       &   in 2036 & in 2037       \\
\hline \multicolumn{10}{c}{solar system dynamical model without four most massive asteroids} \\
 A & 2004\,03\,15 -- 2006\,08\,16 & 1964 & 0\farcs 416 & 0\farcs 418  & 1.3    & -0.12      & $6.151\pm 0.155$   &  1.4$ \cdot 10^{-5}$  &  9.6$ \cdot 10^{-5}$  \\
 B & 2004\,03\,15 -- 2006\,08\,16 & 1950 & 0\farcs 399 & 0\farcs 400  & 0.9    & -0.13      & $6.066\pm 0.149$   &  6$ \cdot 10^{-6}$    &  4$ \cdot 10^{-6}$  \\
 C & 2004\,03\,15 -- 2006\,08\,16 & 1424 & 0\farcs 262 & 0\farcs 263  & $-$0.1 & -0.10      & $5.956\pm 0.106$   &  1.3$ \cdot 10^{-5}$  & $\sim 10^{-7}$\\
 D & 2004\,03\,15 -- 2006\,08\,16 & 1980 & 0\farcs 316 & 0\farcs 317  & 0.2    & -0.12      & $6.074\pm 0.083$   & $5 \cdot 10^{-7}$     & $1.9 \cdot 10^{-6}$  \\
 E & 2004\,03\,15 -- 2006\,08\,16 & 1971 & 0\farcs 308 & 0\farcs 309  & $-$0.1 & -0.10      & $6.065\pm 0.081$   & $6 \cdot 10^{-7}$     & $2.0 \cdot 10^{-6}$\\
\hline \multicolumn{10}{c}{including four most massive asteroids} \\
 E$^{\prime}$ & 2004\,03\,15 -- 2006\,08\,16 & 1971 & 0\farcs 308 & 0\farcs 309  & $-$0.1 & -0.10  & $6.064\pm 0.095$   & $ 7 \cdot 10^{-7}$&$ 1.8\cdot 10^{-6}$\\
 \hline
\end{tabular}}}
\end{center}
\end{table}

\begin{table} \caption{Comparison between selection and
weighting methods taken by different group of researchers for 1000
optical observations of Apophis in the time interval 2004\,03\,15 --
2006\,08\,16; in column 5: no? -- assessment based on the large
number of discarded observations and the value of rms. }
\label{tab:others} \vspace{0.10cm}
\begin{center}
\footnotesize{
\begin{tabular}{cccccc} \hline
 Source                 &  Number of  & Percent of   &Number    & weighting &  rms    \\
                        & used optical& of discarded &of used   &  of       &         \\
                        &   obs.      & optical obs. &radar obs.&  obs.     &         \\
\hline
 \citeasnoun{gio:2008}       & 792         & 21\%         & 7        &  no?      &  0\farcs 407   \\
 \citeasnoun{Vinogradova:08} & 956         & 4.5\%        & 7        &  no       &  0\farcs 370   \\
 JPL SBD                & 738         & 26\%         & 7        &  no?      &  0\farcs 352   \\
 NEODyS                 & 995         & 0.5\%        & 7        &  yes      &  0\farcs 302   \\
 \hline
\end{tabular}}
\end{center}
\end{table}

According to the assumption incorporated in the weighting procedure,
the weighted O-C distributions are normal, e.g. values of kurtosis
are close to zero (see Models~D and~E in Table~\ref{tab:cases} and
Fig.~\ref{fig:OCdist}). When all the residuals greater than the
arbitrarily assumed limit of 0.6 arcsec were rejected the Gaussian
O-C distribution was also obtained (Model~C).

Using the both types of procedures, selection without weighting
(Models A--C) and with weighting (Models D and E) we have determined
-- by the least square method -- the best-fitting osculating orbits
(hereafter nominal orbit) that are now used as a basis for our
impact investigation.

\subsection{Nominal orbit comparison with other orbital calculations}

To compare our solutions with the analogous results in the
literature, we additionally determined the nominal orbital elements
of Model~E for the Epoch's given by \citeasnoun{gio:2008},
\citeasnoun{Vinogradova:08} and two well known Web sources: JPL
Small-Body Database and Near Earth Objects - Dynamic Site (NEODyS)
(see Table~\ref{tab:others}). In all these four sources  the $7$
radar measurements have been incorporated into the orbital
determinations.

We concluded that our values of uncertainties are in excellent
agreement with all of them, except for the uncertainty in the
anomaly given by \citeasnoun{gio:2008}. This one is an order of
magnitude smaller than the uncertainties determined by all the
remaining groups including ours. Our values of the nominal orbital
elements are consistent within $3$~sigma with the results obtained
by other groups. Unfortunately, the comparison with the NEODyS
solution lacks the statistical significance, because the orbital
elements at NEODyS Page are provided without the adequate precision.

\section{Cloning of the nominal orbit}\label{cloning}

\begin{figure}
\begin{center}
\includegraphics[width=8.0cm]{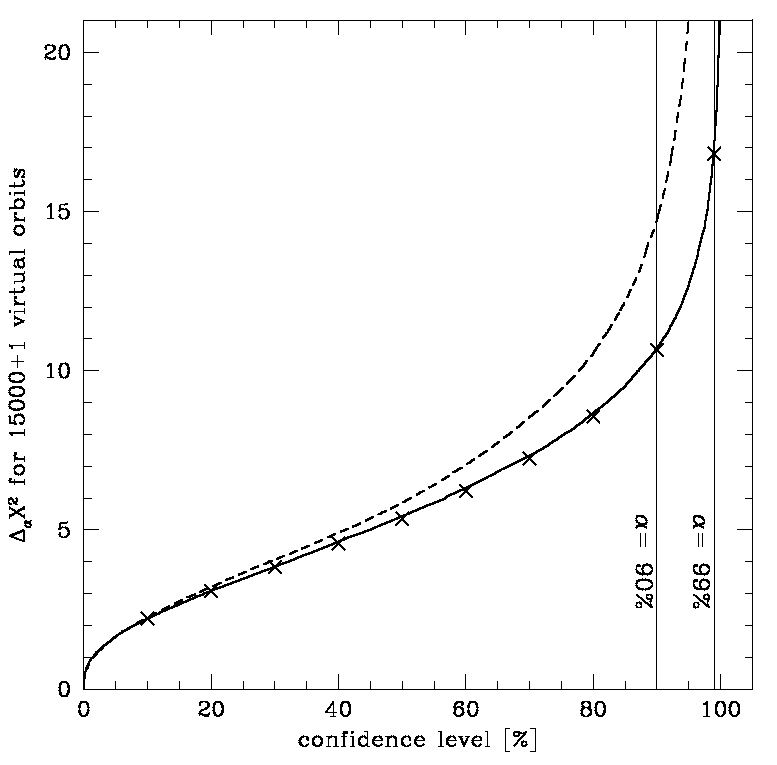}
\caption{ $\Delta _{\alpha} \chi ^2$ for the sample of 15\,000
clones derived in Model~A . Statistics of the sample of cloned
orbits generated at the Epoch of 2006~09~22 is shown with a solid
curve while $\Delta _{\alpha} \chi ^2$ distribution of the sample of
clones generated at the Epoch of 2029~01~24 (three months before
'keyhole' passage) is given in dashed. Grey vertical lines represent
the confidence level of  $90$\,\% and $99$\,\%, respectively.  }
\label{f:chi2}
\end{center}
\end{figure}

\begin{figure}
\begin{center}
\includegraphics[width=13.5cm]{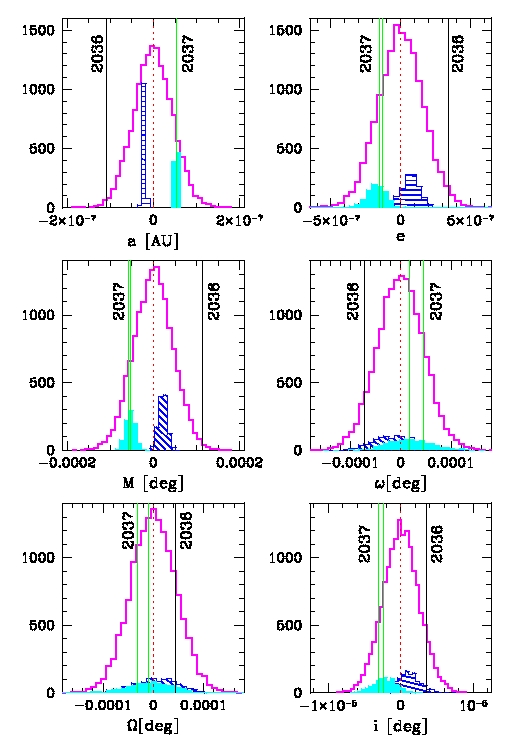}
\caption{The distribution of possible osculating orbits of Apophis
obtained for solution A. The sample of $15\,000$ virtual orbits was
generated for the epoch of 2006~09~22. The plot is centered on the
values of orbital elements of the nominal osculating orbit (epoch:
2006 09 22) represented by dotted vertical lines. Distributions of
VAs which passed closer than $0.04$\,AU in April 2037 (ascending
node) and September 2037 (descending node) are presented by filled
cyan and dashed-filled blue histograms, respectively. The three
impact orbits derived from this sample are shown with the solid
vertical lines (one black impact orbit in 2036, two green -- in
2037).  } \label{f:apo2006}
\end{center}
\end{figure}

\begin{figure}
\begin{center}
\includegraphics[width=9.5cm]{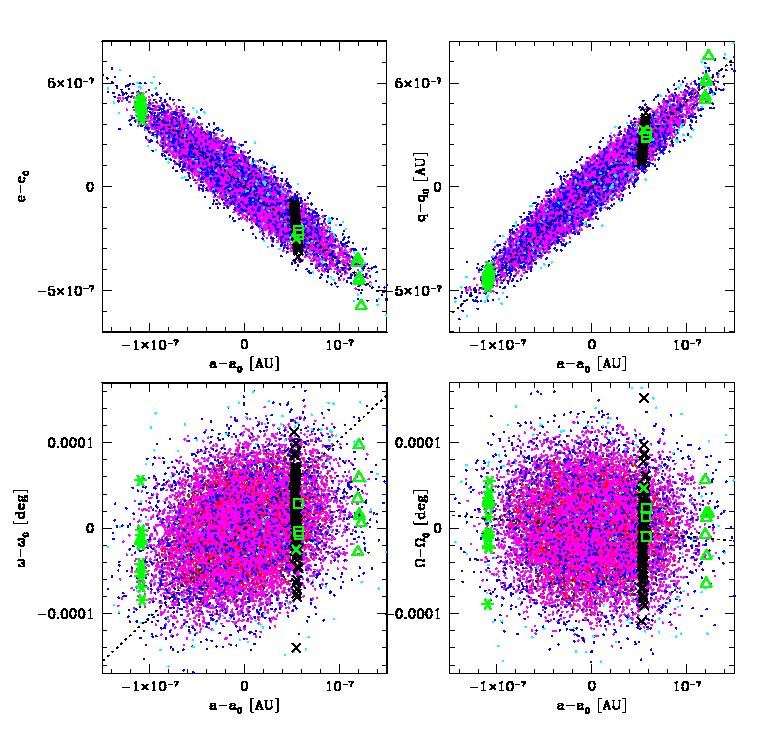}
\includegraphics[width=9.5cm]{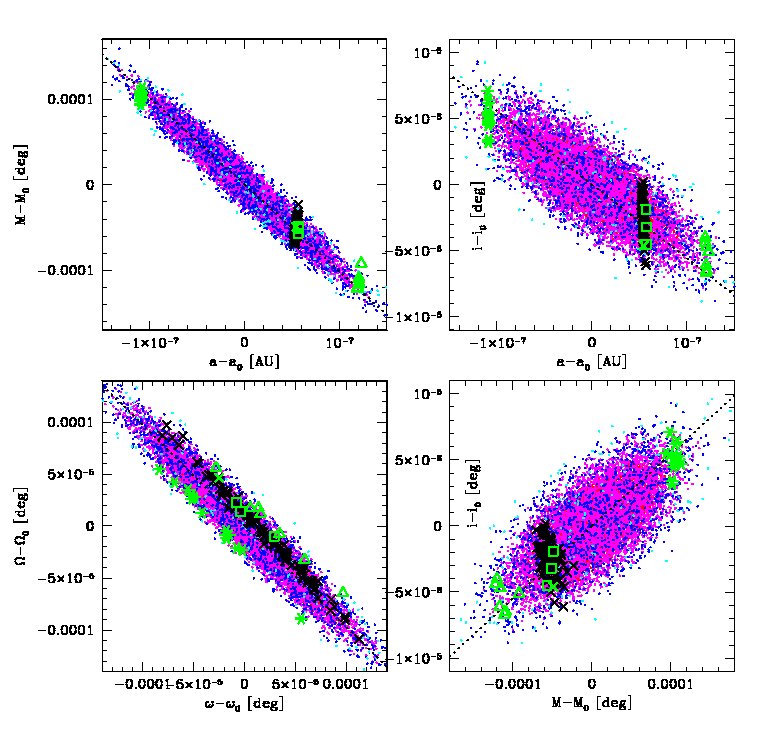}
\caption{Projection of the 6-dimensional space of possible 15\,000
osculating orbits of Apophis onto the plane of two chosen orbital
elements (Solution~A). Each point represents a single virtual orbit,
while the colors indicate the deviation magnitude from the
nominal orbit with the confidence level of: $< 50$\,\%,
$50$\,\%--$90$\,\%, $90$\,\%--$99$\,\%, and $ > 99$\,\% (from the
red to magenta, blue and cyan, respectively). LOV's are given by
black dotted lines. The impact orbits are shown with black crosses
(impact in 2037), green asterisks (impact in 2036) and green triangles
(2046). The derived impact orbits for the year 2054 (green squares)
and 2059 (green cross) are superimposed on the
background of black crosses. Each individual plot is centered on the
nominal values of respective pair of orbital elements denoted by the
subscript '0' (epoch: 2006~09~22). } \label{f:clouds}
\end{center}
\end{figure}

\begin{figure}
\begin{center}
\includegraphics[width=14.0cm]{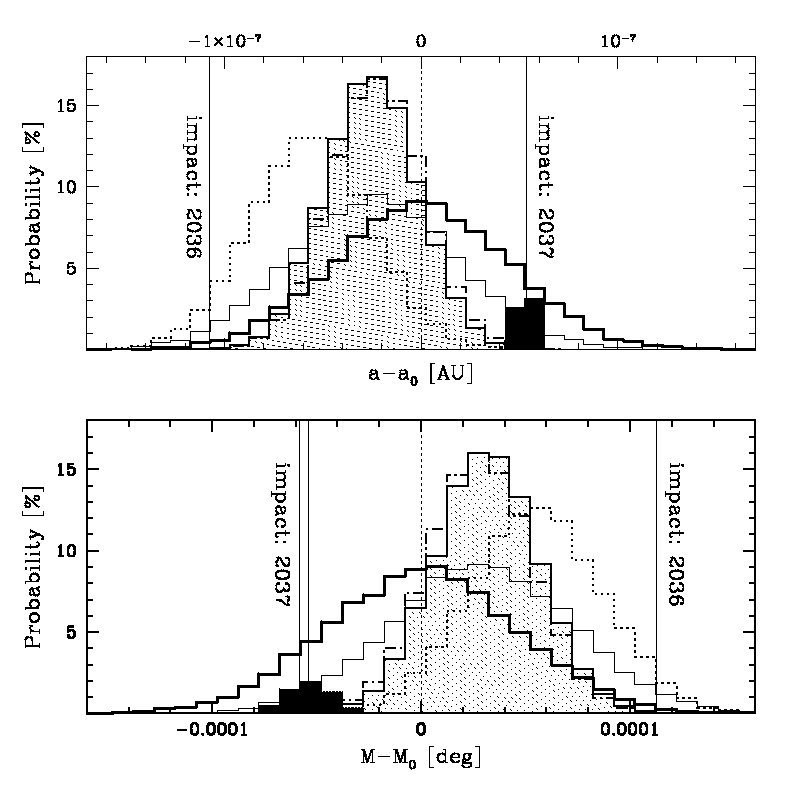}
\caption{ Distributions of the semimajor axes (top panel) and mean
anomalies (bottom panel) derived in models: A -- thick solid line
histogram, B -- thin solid line histogram, C -- dotted line
histogram, D -- dashed dotted line, E -- filled histogram. Each
distribution was constructed on the basis of the samples of
$15\,000$ virtual orbits (Epoch 2006 09 22) and was centered on the
nominal value of semimajor axis (top panel) and mean anomaly (bottom
panel) derived in Model~A. The three impact orbits derived in the
Model~A from the sample of $15\,000$ clones are given as vertical
lines (one impact orbit on April 13, 2036 and two impact orbits --
on April 13, 2037). Distributions of clones which passed closer than
$0.04$\,AU in April 2037 are presented by filled black histograms.}
\label{f:dist5}
\end{center}
\end{figure}

To analyze the impact possibilities in the consecutive encounters of
Apophis with the Earth, it is necessary to examine the evolution of
any possible orbit of Apophis from the confidence region, e.g. the
6-dimensional region of orbital elements where each set of orbital
elements is compatible with the observations. We construct the
confidence region using the Sitarski method of the random orbit
selection \cite{Sit:98}.

Sitarski method allows us to generate any number of randomly
selected orbits of virtual asteroids, (hereafter VAs). The derived
sample of VAs follows the normal distribution in the orbital
elements space. Also the rms's fulfil the 6-dimensional normal
statistics. According to the chi-square test of significance we
have:

\vspace{0.05cm} \noindent $({\rm rms}_i)^2 = ({\rm
rms}_{nom})^2\cdot \left[ 1+ \Delta _{\alpha}\chi
^2/\chi2_{min}\right]$ ~~~~ $i=1,...N$\, ,

\vspace{0.05cm} \noindent where the increment $\Delta _{\alpha}\chi
^2$ is defined by a standard $\chi ^2$ statistics for the the
selected confidence limit, $\alpha$, and the relevant number of
``interesting'' parameters, $N_p$, e.g. number of parameters
estimated simultaneously \cite{Avni:1976}. Since the $\chi ^2$
values are calculated using the sample dispersions, $\sigma$, the
minimum $\chi ^2_{min} = N-N_p$, and in our case $N_p=6$, since six
orbital elements have been simultaneously drawn in the selection of
the cloned orbits.

Critical values of $\Delta _{\alpha}\chi ^2$ one can find in
statistical tables. For example,  for a chi-square distribution with
six interesting parameters we get $90$\,\% of clones with $\Delta
_{\alpha} \chi ^2 < 10.645$ and 99\% of clones with $\Delta
_{\alpha} \chi ^2 < 16.812$. It means that the rms of true (unknown)
orbit of Apophis should satisfy the inequality:

\vspace{0.05cm} \noindent ${\rm rms}_{true} \leq {\rm rms}_{99} =
{\rm rms}_{nom}\cdot \sqrt{1+ 16.812/(N-6)}$ \hspace{0.5cm} for a
confidence level of $99$\,\%. The rms$_{99}$ values are listed in
column~5 of Table~\ref{tab:cases}.

One can see in Fig.~\ref{f:chi2} that orbital cloning procedure at
the epoch relatively close to observational arc provides the
excellent agreement between the derived rms distribution (solid
curve) and the theoretical 6-dimensional normal distribution
(crosses). The same procedure used for the epoch of 2029~01~29 gives
more disperse sample of cloned orbits (the dashed curve) mostly due
to tiny differences in the planetary perturbations for the
individual orbital clones. When the sample of clones selected in
2006 were integrated to the epoch in 2029 the similar dispersion was
observed.

The randomly selected orbits form the confidence region in the
6-dimensional space of possible osculating elements where the
dispersion of the each orbital element is given by its uncertainty
estimated from the least squares method of the orbit determination.

Fig.~\ref{f:apo2006} shows the orbital element distributions of the
sample of 15\,000 VAs for the orbital solution represented by the
nominal orbit of Model~A while Fig.~\ref{f:clouds} presents
projections of the 6-dimensional parameter space of 15\,000 virtual
Apophis onto the plane of two chosen orbital elements. Orbital
cloning procedure was applied at the epoch of 2006 09 22 close to
the observational arc. The derived swarm of VAs follows the normal
distribution in the 6-dimensional space of orbital elements. This is
visualized by four colours of points in Fig.~\ref{f:clouds}. Each
point represents a single virtual orbit, while its colours indicate
the deviation magnitude from the nominal orbit with the confidence
level of: $< 50$\,\%, $50$\,\% -- $90$\,\%, $90$\,\% -- $99$\,\%,
and $> 99$\,\% (from the red points through the magenta, blue and
the cyan points, respectively). The symbols in the crowded areas
heavily overlap and the red points are often covered by the magenta
and blue points.

A comparison between the orbital element distributions in all five
models is given in Fig.~\ref{f:dist5} for the semimajor axis (upper
panel) and the mean anomaly (bottom panel).

\begin{figure}
\begin{center}
\includegraphics[width=8.0cm]{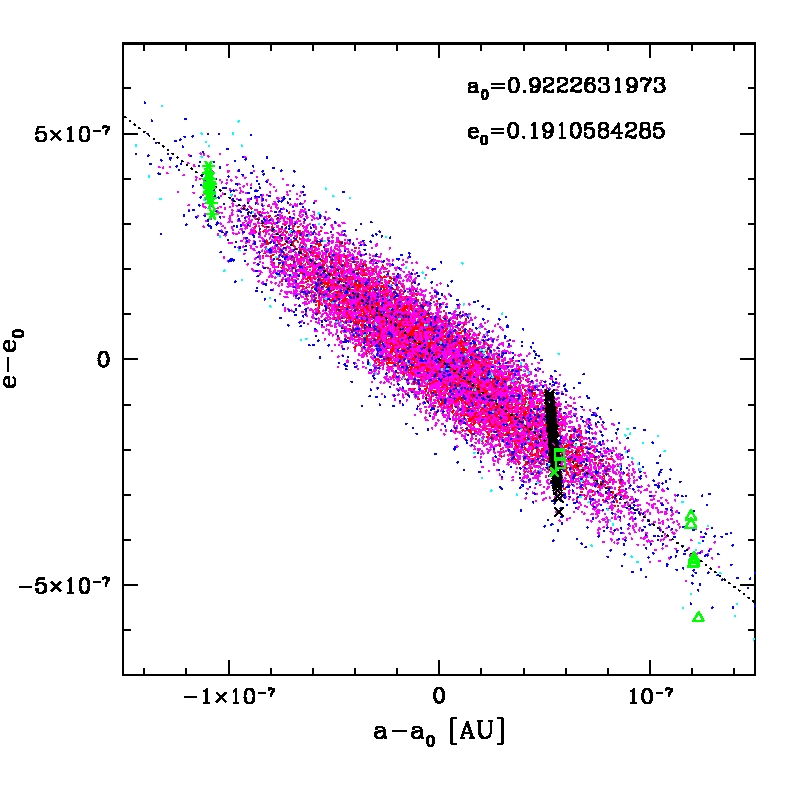}
\includegraphics[width=8.0cm]{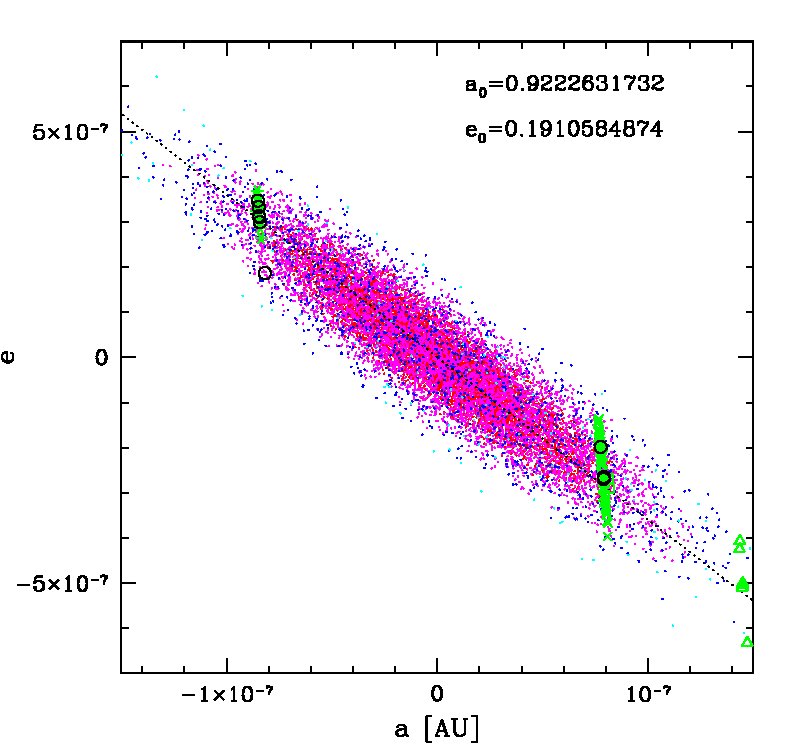}
\includegraphics[width=8.0cm]{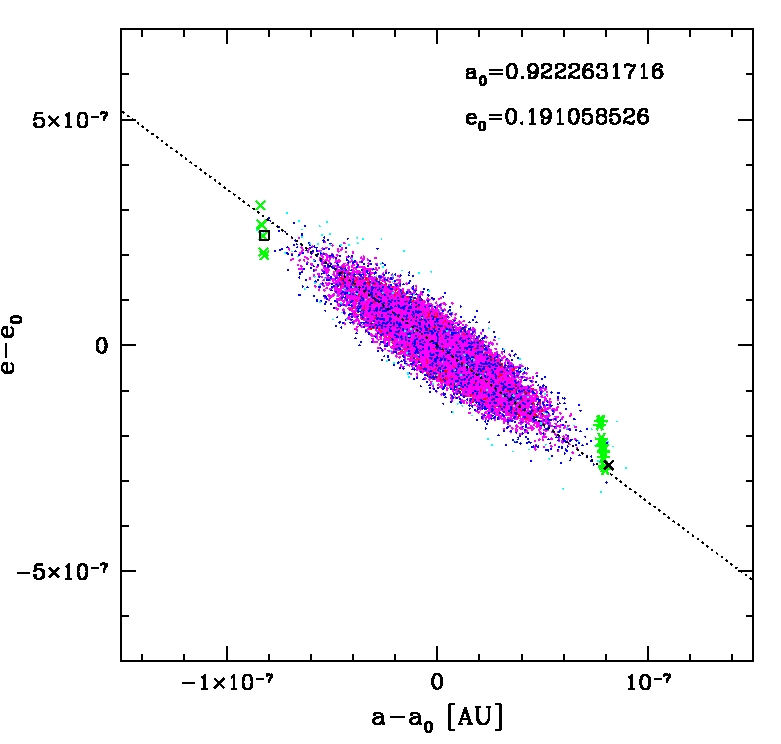}
\caption{ Projection onto the a-e plane in the 6-dimensional space
of possible osculating orbits of Apophis obtained for solution A
(upper left panel), B (upper right panel) and E$^{\prime}$ (lower
panel). Samples of 15\,000 VAs are given by points coded as in
Fig.~\ref{f:clouds}. All derived impact orbits in Model~A (from 1
million VAs) are shown in left upper panel as the same symbols as in
Fig.~\ref{f:clouds}. The impact VAs derived in Model~B (impacts in
2036 and 2037 detected from 1 million VAs) are shown with the
black open circles in the right upper panel (on the background of
green impact VAs from Model~A), while the impact orbits derived in Model
E$^{\prime}$ (impacts from 10 million VAs) are shown in the bottom
panel (black cross -- one impact VA in 2054 on the background of 18
impact VAs in 2036, black square -- one impact VA in 2044 at the
background of 7 impact VAs in 2036). Each plot was centered on the
values of the semimajor axis, $a_0$, and eccentricity, $e_0$, of the
nominal orbit given in the right corners of each plot (epoch: 2006
09 22). } \label{f:ae}
\end{center}
\end{figure}

\section{Impact analysis of  Apophis: the method and results}\label{impactan}

We are able to directly determine the sample of impact orbits for
the each close encounter with the Earth whenever such risk orbits
exists \cite{Sit:02}. However, for the analysis of the impact
probability we have developed a new method.

To examine the Apophis close encounter with the Earth in 2029 and
impact risk in the following years, the non-linear two-stage
analysis was performed numerically.

In the first step, we constructed the sample of $15\,000$ clones
($15\,000$ of VA's) (see Sect.~\ref{cloning}) for each of the
orbital solutions described in Table~\ref{tab:cases}. Each of these
orbital clones was then integrated forward in time to the year 2100.
Thus, we integrate the swarm of virtual asteroids (VAs) from the
whole uncertainty region, not only VAs lying on the line of
variations (LOV). Additionally, our swarm of VAs follows the normal
distribution in the orbital elements space.

From these $90\,000$ VAs we have obtained three impact orbits in the
Model~A (one in 2036 and two in 2037) and one impact orbit for the
Model~B (in 2036). We also noticed that about $6-7$\,\% of VAs
(depending on the model) passed Earth on April 2051 at the distance
within $\sim 0.000069 - 0.04$~AU. Vertical lines in
Fig.~\ref{f:apo2006} present the positions of impact orbits in the
sample of $15\,000$ clones in Model~A, while the positions of impact
orbits in the semimajor distributions and mean anomaly distributions
for all five models are shown in Fig.~\ref{f:dist5}. One can see
from Fig.~\ref{f:apo2006} that the range of semimajor axes including
all the clones passing in 2037 closer than $0.04$\,AU from the
geocenter (filled histogram) is relatively narrow in comparison with
the full $a$-distribution. Analogous ranges in the remaining orbital
elements are more disperse.

In the second step, on the basis of obtained impact orbits (four
orbits in this example), we construct the potentially 'dangerous'
intervals of semimajor axes at our epoch of orbital cloning
(2006~09~22). The 'dangerous' ranges of semimajor axes were also
independently derived for all dates of potential impacts or close
encounters using Sitarski method (\citeasnoun{Sit:06}). In this way we are
able to randomly select large number of VAs (for each model of data
selection and weighting) and then take for the numerical integration
only VAs within 'dangerous' interval for the given moment of impact.
After many tests it turned out that these 'dangerous' intervals are
very narrow in the $a$-distribution. Thus, to evaluate the true
probability of the impact, it was possible to randomly select
million of clones and then effectively integrate only a thousands or
dozens of thousands of clones. It is important to stress here that
the impact probabilities given in Table~\ref{tab:cases} were always
estimated from the samples of at least one million of randomly
selected VAs. Finally, we detected 96 impact orbits in April 2037
and 14 impact orbits in April 2036 in the Model~A (non-weighted
observations, sample of one million VAs). We also detected impact
orbits in 2036 in the Model~B (6 events) and in Model~C ($13$
events) while in Models~D and~E we have got no impact orbit in 2036
from the sample of one million randomly selected orbits. However, in
the sample of $10$ million VAs, five and six impact VAs in 2036 were
detected (in models~D and E, respectively), and $19$ and $20$ impact
VAs in 2037.

These results can be qualitatively explained by the positions of
impact orbits relative to the $a$-distribution in the top panel of
Fig.~\ref{f:dist5}. The semimajor axis $a_0$ of the nominal orbit of
the Model~A has been selected as a reference value at the abscissa
for all five models. For this reason, the histograms for the
Models~B through E are displaced from the central position. Solid
vertical lines indicate semimajor axes of the impact orbits and one
can see why for the weighted observations (Models~D and E) the
impact in 2036 and 2037 has a significantly lower probability than
for the Models~A and B, (obviously, the histograms for $15\,000$ VAs
are not representative for wings of the distributions of $1-10$
million VAs).

Additionally, the probability of about $8\cdot 10^{-6}$ for the
impact in 2046 was estimated in the Model~A (8 impact orbits from
one million of clones). On the basis of these eight impact orbits we
have calculated the keyhole of $2.9$\,kilometer wide at a distance
of $6.5702-6.5706\,R_{\oplus}$ from the Earth's center on April~13,
2029 (Table~\ref{tab:keyholes}).

A careful analysis of the VAs orbits has revealed several new
interesting results. When we examined the 'dangerous' interval for
the impact risk in 2037, we have detected a series of impacts in the
years following the year 2037. Firstly, we have found two new
keyholes on April~13, 2029 -- closely related to the 2037 keyhole:
the keyhole of $\sim 1.3$\,kilometer wide lying at a distance of
$6.3486-6.3488\,R_{\oplus}$ from the Earth's center (calculated from
3 impact orbits in 2054), and the keyhole at a geocenter distance of
$\sim 6.3359\,R_{\oplus}$ (estimated from one impact orbit in 2059).
Secondly, we derived very special impact orbit in 2076 connected
with very close encounter with Earth in 2051. In our basic samples,
as was mentioned before, about $6-7$\,\% of VAs (depending on the
model) passed Earth at the distance of $\sim 0.000069 - 0.04$\,AU.
However, only these VAs that in 2051 pass near the Earth almost
exactly at the distance of $\sim 0.00819$\,AU have a chance to hit
the Earth in 2076. Since the keyhole in 2029 is extremely narrow for
the impact in 2076, the probability of this impact is lower than the
probability of impact in 2036 though the VAs hitting the Earth in
2076 are much closer to the nominal orbit than the VAs impacting on
the Earth in 2036.

Detected impact orbits from the swarm of one million VAs are shown
in Fig.~\ref{f:clouds} superimposed on the sample of $15\,000$ VAs
constructed for the Model~A in the first step of our analysis. One
can see that each projection of the impact VAs onto a plane of a
pair of orbital elements forms elongated structure for a given
impact date. One should note that these structures generally (though
not always) intersect the LOV projection (Fig.~\ref{f:clouds}).
Thus, if the search is limited only to this line, some impact orbits
would be found. Nevertheless, most of the impact orbits are situated
far from the LOV and to find all the possibilities of impact orbits
one should examine the entire 6D volume of the orbital element
space.

The comparison between the swarms of 15\,000~VAs derived in
models~A, B and E$^{\prime}$ and the impact VAs in these three
Models detected in $1-10$ million VAs are shown in Fig.~\ref{f:ae}.

\subsection{Trajectory prediction uncertainty at the moment of close
encounter in 2029}\label{encounter2029}

\begin{figure}
\begin{center}
\includegraphics[width=14.0cm]{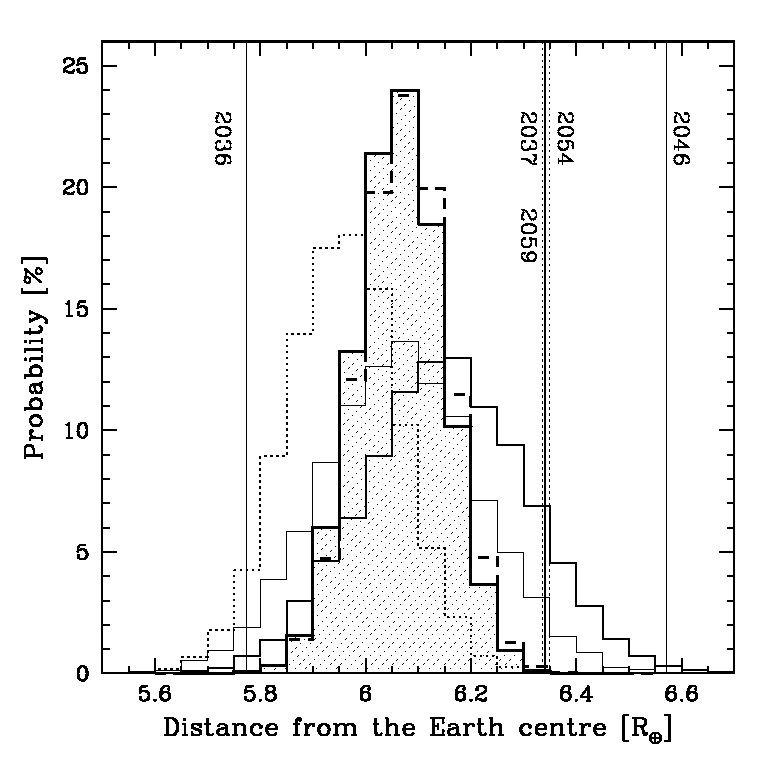}
\caption{ Distributions of the minimum distance of the asteroid
Apophis from the centre of Earth in 2029 04 13 derived for the
samples of $15\,000$ virtual orbits. The minimum distance histograms
for the Model~A is shown with a thick solid line, for the Model~B --
thin solid line, Model~C -- thin dotted line, Model~D -- thick
dashed histogram. For the most accurate Model~E the distribution is
shown with the thick solid line and filled histogram. The distances
for the selected impacts at dates indicated by labels is shown.}
\label{f:keyhole}
\end{center}
\end{figure}

The probability distributions of the distance encounter with the
Earth on April 13, 2029 are shown in Fig.~\ref{f:keyhole} for all
five models of the data processing. Each histogram was constructed
for 15\,000 VAs. The expected values of the Apophis distance
encounter with the Earth are calculated by fitting the normal
distribution to each of these histograms. The results are given in
column~8 of Table~\ref{tab:cases}. Weighted mean value of  the
geocentric encounter distance calculated from all five minimal
distance estimations is equal to $6.055\pm 0.099$\,$R_{\oplus}$.

One can see that including four most massive asteroids into the
solar system dynamical model makes the distribution of the minimal
distance during close encounter event in 2029 wider by about
$17$\,\% (compare Model~E and Model~E$^{\prime}$; the selection and
weighting of data are the same in both models). According to
\citeasnoun{gio:2008} these four asteroids constitute about $68$\,\%
of the whole asteroid perturbers during 2004-2036. Thus, we expect
that this minimal distance distribution in 2029 becomes wider by
about next $8$\,\% with  $\sigma \simeq 0.101$\,R$_\oplus$ in the
case of Model~E. However, we have obtained quite similar
probabilities for the impact in 2036 ($6\cdot 10^{-7}$ and $7\cdot
10^{-7}$) and 2037 ($2.0\cdot 10^{-6}$ and $1.8\cdot 10^{-6}$) in
both models (E and~E$^\prime$). Once again, specific features of the
investigated events indicate significance of the nonlinear effects
in the impact analysis of Apophis. For example, in the Model~E we
derive one impact VA in 2044 (from the sample of 10 millions of VAs;
not shown in Fig.~\ref{f:keyhole}) that previously passed close to
the Earth in 2037 whereas in Model~E$^\prime$ we also derived one
impact VA in 2044, however this VA passed near the Earth in 2036. It
was found that the first impact clone was placed in the right wing
of the 2029 keyhole distribution in Model~E (Fig.~\ref{f:keyhole})
while the second -- in the left wing of the keyhole distribution in
Model~E$^\prime$ (notice that both minimal distance distributions
are centered on the same value of 6.06\,R$_{\oplus}$).

Table~\ref{tab:cases} and Fig.~\ref{f:keyhole} show that models
based on the weighted observations are most accurate and very
similar. The best solutions give geocentric encounter distance of
$6.065\pm 0.081$\,$R_{\oplus}$ (Model~E) or $6.064\pm
0.095$\,$R_{\oplus}$ (Model~E$^{\prime}$) on April13, 2029. Both
values are in excellent agreement with $5.96\pm 0.09$\,$R_{\oplus}$
given by \cite{gio:2008} as the best estimate of the geocentric
encounter predicted from the optical observations  {\it and} the
radar measurements. One should note that uncertainties of the
predicted close encounter in 2029 derived in our
Model~E/E$^{\prime}$ have the same accuracy as solution  which
include to the orbital fitting also the radar observations.

Table~\ref{tab:keyholes} presents the range of Earth's distances of
all the numerically detected impact keyholes at the moment of the
close encounter with the Earth on April 13, 2029. These keyholes
that were detected in Model~A from one million of VAs are shown by
vertical lines in Fig.~\ref{f:keyhole}. It is important to stress
that the symbols $< 10^{-6}$ (or $< 10^{-7}$) given in
Table~\ref{tab:keyholes} only inform that we did not found any
impacts in one million ($10$ million) of VAs. One should notice that
in the case of the non-weighted data (Models~A--C) we perform
analysis based on one million VAs whereas for the weighted models
(Models~D, E and E$^{\prime}$)-- on $10$ million VAs.

Including the four most massive asteroids into the solar system
dynamical model does not affect significantly the position of the
2029~04~13 keyholes for the impacts in 2036, 2037 and 2046. However,
the impacts in all the remaining years listed in the
Table~\ref{tab:keyholes} followed after the close encounter of the
VA with the Earth. Therefore, the evolution of such VAs is very
sensitive even to the very small additional perturbations, including
the perturbations from the massive asteroids. An example of such
perturbation was discussed in this section in the context of the
impact orbit in 2044.

\begin{table}
\caption{Keyholes for the potential impacts in 2036, 2037, 2046 and
impacts in 2044, 2054, 2055, 2056, 2059, 2076 that are preceded by a
close encounter with Earth in 2036, 2037, 2046 or 2051}
\label{tab:keyholes} \vspace{0.10cm}
\begin{center}
\footnotesize{
\begin{tabular}{ccrccccc} \hline
 Potential        &   Keyhole at the Epoch& \multicolumn{5}{c}{Impact}      \\
 impact in        &   of 2029 04 13       & \multicolumn{5}{c}{probability} \\
   April:         &   [R$_{\odot}$]       &      Model A             &    Model B          &    Model C          &    Model D &    Model E   \\
\hline
 2036             & 5.7736--5.7744        &  1.4$\cdot 10^{-5}$      &  0.6$\cdot 10^{-5}$ &  1.3$\cdot 10^{-5}$ & $ 5 \cdot 10^{-7}$  & $6 \cdot 10^{-7}$   \\
 2053             & 5.7763                &  $ < 10^{-6}$            &  $ < 10^{-6}$       &  $ < 10^{-6}$       & $ \sim 10^{-7}$     & $< 10^{-7}$ \\
&&&&&& \\
 2076             & 5.97347               &  $\sim 10^{-6}$          &  $<10^{-6}$         &  $<10^{-6}$         & $< 10^{-7}$         & $\sim 10^{-7}$ \\
&&&&&& \\
 2059             & $\sim$6.3359          &    $\sim  10^{-6}$       &  $<10^{-6}$         & $<10^{-6}$          & $<10^{-7}$          & $<10^{-7}$  \\
 2044             & $\sim$6.3370          &    $< 10^{-6}$           &  $<10^{-6}$         & $<10^{-6}$          & $\sim 10^{-7}$      & $\sim 10^{-7}$  \\
 2037             & 6.3395--6.3405        &  9.6$\cdot 10^{-5}$      &  4$\cdot 10^{-6}$   & $\sim 10^{-7}$      & $1.9 \cdot 10^{-6}$ & $2.0 \cdot 10^{-6}$  \\
 2056             & 6.3426                &    $< 10^{-6}$           &  $<10^{-6}$         & $<10^{-6}$          & $<10^{-7}$          & $<10^{-7}$  \\
 2054             & 6.3486--6.3488        &  3$\cdot 10^{-6}$        &  $<10^{-6}$         & $<10^{-6}$          & $3 \cdot 10^{-7}$   & $\sim 10^{-7}$  \\
&&&&&& \\
 2046             & 6.5702--6.5706        &  8$\cdot 10^{-6}$        &  $<10^{-6}$         & $\ll 10^{-6}$       & $\ll 10^{-7}$       & $\ll 10^{-7}$  \\
 2055             & $\sim$6.5739          &  2$\cdot 10^{-6}$        &  $<10^{-6}$         & $\ll 10^{-6}$       & $\ll 10^{-7}$       & $\ll 10^{-7}$  \\
 \hline
\end{tabular}}
\end{center}
\end{table}

\subsection{Impact orbits far from the nominal orbit of Apophis}

Analyzing the impact possibilities based on the shorter arcs of
observations (see arc3 and arc5 in Table~\ref{tab:arcs}) we have
derived many impact orbits in 2048 and many other impact
possibilities which would take place in the following years (2049,
2062, 2063, 2065) that are connected with close encounter in 2048.
According to the present interval of observations all these impact
events are practically excluded since they would take place if
Apophis passes near the Earth in 2029 at the distance of 7.060 --
7.063\,R$_{\oplus}$. One can see in Fig.~\ref{f:keyhole} that this
keyhole for impact in 2048 is far on of the left wing of displayed
distributions for the current orbital models of Apophis. Still
further on the left wing in Fig.~\ref{f:keyhole}  are distributed
the keyholes for impacts in 2053 and 2067 discussed by
\citeasnoun{Sit:06} on the basis of the non-weighted data.

\subsection{Orbital evolution of Apophis after 2029}

During the incoming first close encounter with the Earth on April
13, 2029 the orbit of Apophis will change. The most significant
change from $0.92$\,AU to $1.10$\,AU will affect the semimajor axis.

\begin{figure}
\begin{center}
\includegraphics[width=16.0cm]{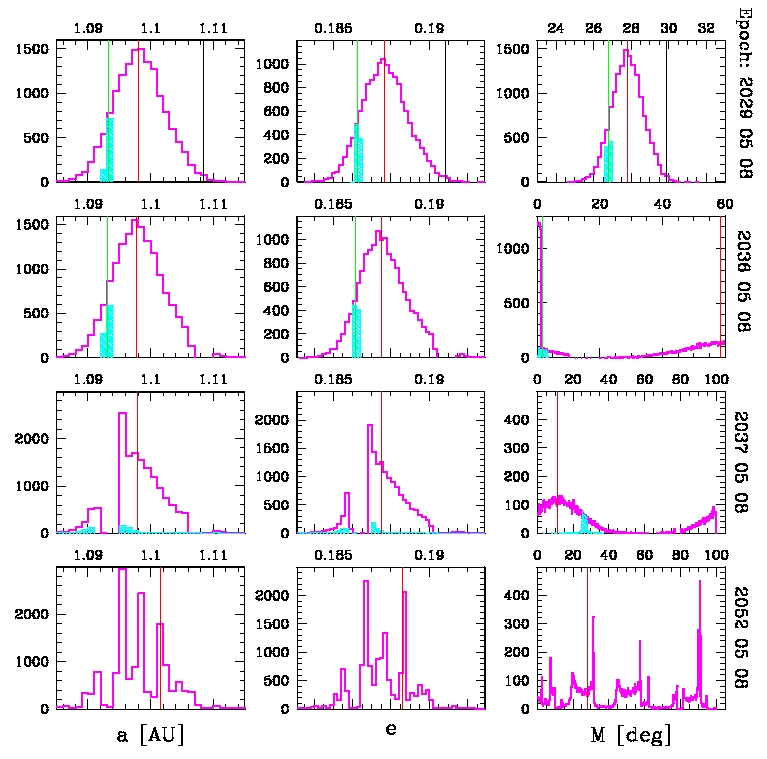}
\caption{ Evolution of potentially possible osculating orbits of
Apophis obtained for solution A. The starting sample of $15\,000$
virtual orbits was taken at the epoch of 2006 09 22. Time runs from
top to bottom and the epoch of displayed distributions are given at
the right-hand side of each row. The top row represents the
distributions of a, e and M about one month after very close
encounter with Earth in April 2029. The position of the evolved
nominal orbit is shown with the red vertical line. Distributions
of VAs which passed closer than $0.04$\,AU in April 2037 are
presented by filled cyan histograms. Three impact orbits derived
from this sample are given as black vertical line (one impact orbit
on April, 13 2036) and green vertical lines (two impact orbits on
April 13, 2037). }
\label{fig:ewo1}
\end{center}
\end{figure}

\begin{figure}
\begin{center}
\includegraphics[width=16.0cm]{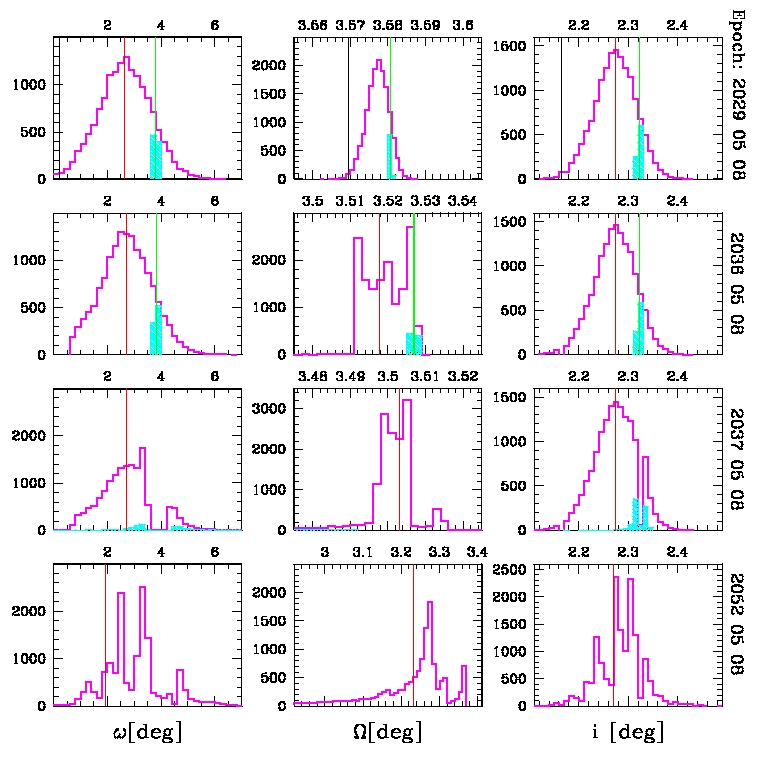}
\caption{ Same as Fig.~\ref{fig:ewo1} for $\omega$, $\Omega$ and
orbital inclination $i$. } \label{fig:ewo2}
\end{center}
\end{figure}

In the top row in Figs~\ref{fig:ewo1},~\ref{fig:ewo2} the
distributions of six orbital elements at the epoch of 2029~05~08 are
shown for Model~A. Apparently, after the close encounter on April
13, 2029 the distributions of parameters of the clone swarms are
still close to the normal distributions, although with several
orders of magnitude greater dispersions than those for the swarm
drawn for the epoch 2006~09~22 (Fig~\ref{f:apo2006},~\ref{f:dist5}),
or any epoch before the close encounter in April 2029. Comparing the
dispersions of semimajor axes (perihelion distance) we have found
that the dispersion increases five orders of magnitude from $\sim
4.5\cdot 10^{-8}$\,AU$\simeq 6.7$\,km ($\sim 2\cdot
10^{-8}$\,AU$\simeq 3$\,km) at the epoch of 2029~01~24 to $\sim
5\cdot 10^{-3}$\,AU ($\sim 2.5\cdot 10^{-3}$\,AU) at the epoch of
2029~05~08. Generally, an ellipsoid of the orbit uncertainty grows
in each orbital element at least four orders of magnitude due to the
close encounter with the Earth on April 13, 2029.

One can see that our Model~C with the smallest rms gives exactly the
same minimal distance of the nominal orbit on April 29, 2029 as
\citeasnoun{gio:2008} (see Table~\ref{tab:cases}).  However, in
Model~C no weighting was applied and almost $29$\,\% of the optical
data were discarded. Because we believe that such extensive
rejections of the modern data are unjustified and inappropriate, we
consider our Model~E as the best in the statistical sense.  We found
that in Model~E the uncertainty along the orbit path on April 13,
2036 is analogous to those presented in Fig.~4 of
\citeasnoun{gio:2008}(10000 VAs) who included the 7 radar
measurements into the orbit determination and discarded about 20\%
of the optical data.

After the close encounter in 2029 the distributions of orbital
elements are not adequately described by the normal distributions.
Consecutive returns to the Earth significantly change the orbital
elements of these clones that pass through the small keyholes in
2036 and 2037. It is demonstrated by the distribution of the
semimajor axes and the eccentricities for Model~A (the first and
second column in Fig.~\ref{fig:ewo1}, respectively). In the first
row positions of the impact clones in April 2036 and in April 2037
(solid vertical lines) are shown. The subsample of clones that
passes closer than $0.04$\,AU in April 2037 are presented by filled
histograms. In May 2036 (second row in Fig.~\ref{fig:ewo1}) one can
see a deficit in the Gaussian shape around the position of the
impact orbit in 2036. This results from the Earth's perturbations
which have changed significantly orbits of these clones. Second and
very prominent dip appears in May 2037 on the position of the
subsample of clones that passed in April 2037 closest to the Earth.
One can see that these clones were almost completely removed from
the narrow interval of semimajor axis and eccentricity and were
dispersed over the rest of the histogram (filled distributions in
the third row in Fig.~\ref{fig:ewo1}). Distributions derived in May
2052 display many similar dips that were created by the periodic
relatively strong Earth's perturbations (which would take place
between 2037 and 2052) that affect different parts of distributions.

\section{Summary and conclusions}

\vspace{0.5cm}

Though Apophis motion is well predictable before its deep close
encounter with the Earth on April 13, 2029, the present observations
are not adequate to eliminate definitely the possibility of impact
with the Earth in 2036 and in many years following this year even in
fully ballistic model. It was shown that the available seven radar
measurements are not crucial at present for the nominal orbit
determination, though historically were important for indication
that the prediscovery observations of March 2004 were biased by some
systematic errors. In the present paper we thoughtfully inspected
the observational material. The data used in the calculations have
been selected according to the well defined and objective
statistical criteria. Our best solution for the passage on April 13,
2029 give the geocentric encounter distance of $6.065\pm
0.081$\,$R_{\oplus}$ (without perturbations from asteroids, Model~E)
or $6.064\pm 0.095$\,$R_{\oplus}$ (including perturbations from four
largest asteroids, Model~E$^{\prime}$). Both values are in excellent
agreement with the results by \citeasnoun{gio:2008} which
incorporated also the radar measurements.

We carefully examined the Apophis impact possibilities with the
Earth after 2029 for VAs that will pass near the Earth at the
distance between $5.6$\,R$_{\oplus}$ and $6.6$\,R$_{\oplus}$ on
April 13, 2029. We show that the impact keyholes in 2036 and 2037
(or group of impact keyholes connected with the close encounter with
the Earth in 2036 and 2037) are placed on the opposite wings of the
normal distribution of the minimal distance in 2029.

Our calculations provide different sizes of the keyholes from those
available in the literature because our impact analysis is based on
the VAs which fill the entire volume of 6D space, while the other
impact results are limited -- as far as we know -- just to the line
of variations (LOV) in the parameter space. We show explicitly that
some of the potential impact orbits do not lie on the LOV.

These two keyholes (or two keyhole groups) listed in
Tab.~\ref{tab:keyholes} are separated by about $0.56$\,R$_{\oplus}$.
Our best Model~E/E$^{\prime}$ are placed almost exactly in the
middle between these impact keyholes. This geometry is very
fortunate from the point of view of the impact risk, assuming than
no other impact keyhole exists within this region. Unfortunately
between them we detected narrow impact keyhole for the collision in
2076. This keyhole is situated extremely close to the nominal orbit
determined by \citeasnoun{gio:2008} -- it is separated only by about
$0.01$\,R$_{\oplus}$ from their nominal value, while the nominal
orbits derived in Model~E/E$^{\prime}$ differ by about
$0.09$\,R$_{\oplus}$ ($\sim$\,one sigma) from impact keyhole for
2076 collision. The \citeasnoun{gio:2008} value is separated
$0.19$\,R$_{\oplus}$ ($\sim$\,two sigma), and $0.38$\,R$_{\oplus}$
($\sim$\,three sigma) from the impact keyholes in 2036 and 2037,
respectively. It will be important to take all these detected
keyholes into considerations during the planned mission of Foresight
spacecraft or any other mission to Apophis.

\section*{Acknowledgments}
This work was partly supported by the Polish Committee for
Scientific Research (the KBN grant 4~T12E~039~28).


\bibliography{comets3}


\end{document}